\documentclass[twocolumn,twoside,slac]{revtex4}
\usepackage{graphicx}
\usepackage{fancyhdr}
\begin{document}

\title{Online and Offline Computing systems in the PHENIX experiment}

%

\author{Martin L. Purschke, for the PHENIX collaboration}
\affiliation{Brookhaven National Laboratory, Upton, NY 11973, USA}

\begin{abstract}

PHENIX~\cite{phenix} is one of two large experiments at Brookhaven
National Laboratory's Relativistic Heavy-Ion Collider (RHIC). At the
time of the conference, the PHENIX experiment was about halfway
through the 2003 run, which started in January 2003. In preparation
for the run, the PHENIX data acquisition, the computing
infrastructure, and the software have undergone several upgrades.
Those upgrades boost the recorded data rate to about 100\,MB/s and allow
for a fast reconstruction only a few weeks after the data have been
taken. As part of the upgrade, essentially all servers in the Online
System have been converted from Solaris to Linux, and a new Linux
computing farm has been commissioned at the experimental site that is
used to prepare for a rapid offline reconstruction pass. This paper
presents a general overview of PHENIX computing. We will explain the
current status, the changes, choices of software and hardware, and
discuss our experience with the new setup.
\end{abstract}

\maketitle

\thispagestyle{fancy}


\section{Introduction}

PHENIX [1] consists of 4 large spectrometer arms, two central arms and
two forward Muon arms. There is a total number of about 350,000
readout channels, giving a typical event size of about
110\,KB/event. The typical sustained readout rate is slightly more than
1\,KHz, which results in a data rate of about 120\,MB/s. The so-called
Run~3 of RHIC was set up with d-Au beams at 200\,GeV/c per nucleon, and,
in the second half, polarized protons at 200\,GeV/c.

In the months leading up to Run~3, which started in January 2003,
we upgraded several key components of the data acquisition and the online
computing system. 
\begin{itemize}
\item
we replaced essentially all servers running the Solaris operating
system with machines running Linux;
\item
we replaced a component in the data transfer, the ``ET'' system, 
with a server process that significantly reduces overhead and allows a data logging rate of 
about 120\,MB/s.
\item
we commissioned another computing farm at the experimental site that is used for
online calibration-related analysis and event filtering.
\end{itemize}

\section{Data Flow in PHENIX}

\begin{figure*}
\centering
\includegraphics[width=125mm]{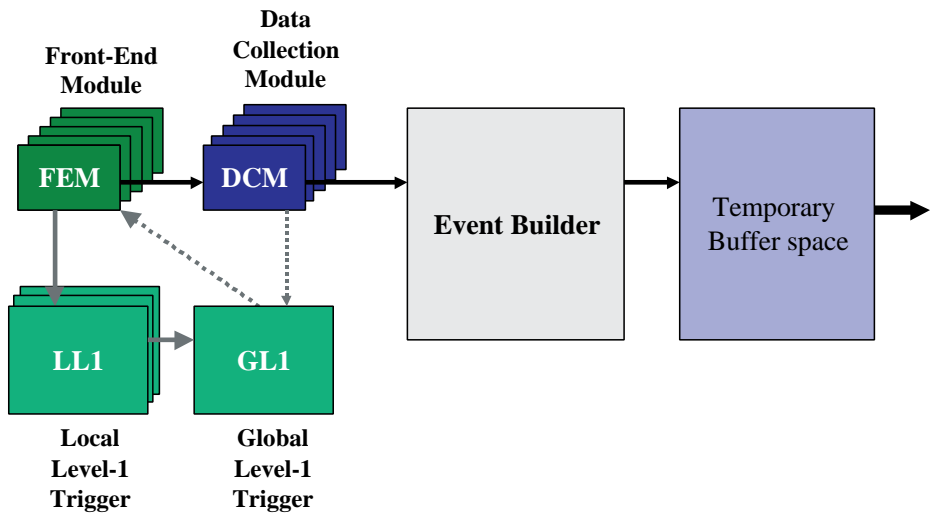}
\caption{\label{dataflow1}
Schematic view of the data flow in PHENIX.}
\end{figure*}

\begin{figure*}
\centering
\includegraphics[width=125mm]{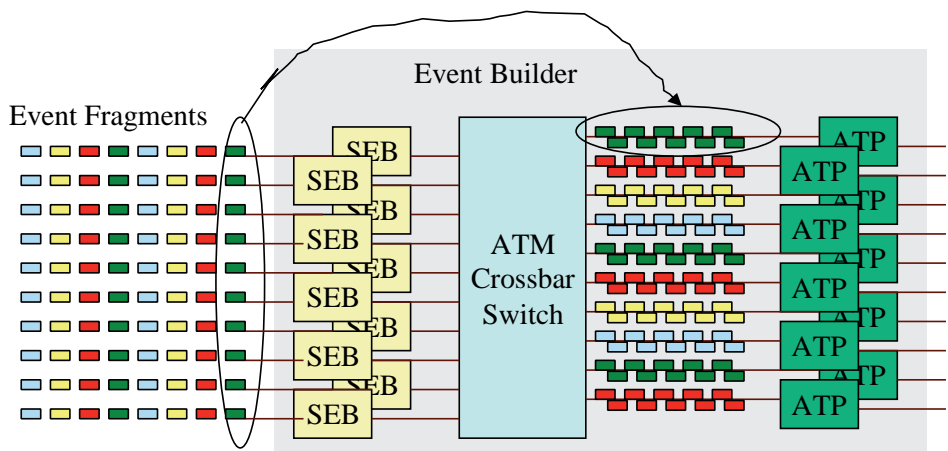}
\caption{\label{evb}
Schematic view of Event builder. The SEBs receive the data 
from the DCMs and send the
parts from one event through a crossbar switch to an ATP, which assembles the event
and runs a Level2-algorithm of the event data.}
\end{figure*}

Fig.~\ref{dataflow1} shows a schematic view of the data flow in PHENIX. 
The detector signals are digitized in Front-End Modules
(FEM) with a number of channels ranging between 12 and several
hundred, depending on the detector. The digitized data are sent via
optical fiber to Data Collection Modules (DCM), which receive the
data, package them, and send the data on to the Event Builder, which assembles 
the event fragments into whole events (fig.~\ref{evb}).

The SubEvent Buffers (SEB) receive the data from the DCMs and send the
parts from one event through an ATM-based crossbar switch to an
Assembly and Trigger Processor (ATP). While the number of SEBs is
fixed by the way the front-end connections are structured, the number
of ATPs is not fixed and more machines can be added to increase the
computing power. The ATPs assemble the event fragments into a whole
event and, because this is the first time that the data of a whole
event are available in one place, runs a Level-2 trigger on the data.

The ATPs pass the data on to ``Buffer Boxes'', two PC's with 2 TB disk
space each, which temporarily store the data on disk until they are
shipped off to an HPSS-based tape robot in the RHIC Computing Facility
(RCF). Those PC's have dual Alteon Gigabit network interface cards, one
on the local network and the other one on a high-speed network connecting
to the RCF. The latter network uses Jumbo frames (9000 bytes MTU size).

The buffering of the data on the buffer disks allows the sending of
data to the RCF to gracefully survive any short-term service
interruptions of the tape robot. This buffering also enables us to
feed the tape robot a steady stream of data independent of the ebb and
flow of the DAQ system. And since the transfer to HPSS does not need
all the capacity of the buffer machines, the data are available for
near-line analysis processes such as detector calibrations and event
filtering for several hours before we need to make room for new data.

The data logging mechanism was replaced recently. Before, we used a
system called ``ET'' (``Event Transfer'')~\cite{timmer} (explained in
more detail later) to merge the data streams from the
ATPs. Events get injected into an ET system and retrieved by the
logger and written to disk. For efficient network transfers, the ATPs
send buffers of data of variable length with an average number of
about 15 events. In the old system, this buffer would get disassembled
into individual events which got injected into the pool, then
re-assembled into buffers by the logger. This caused a lot of overhead
that limited the data rate to about 25\,MB/s per buffer machine. 

We replaced this system with the \emph{Advanced Multi-threaded Logger}
(AML). The AML spawns a thread for each incoming network connection
from the ATPs, and the threads receive the data from the network in
parallel. When a buffer is received, the thread obtains a lock on the
output file and writes the buffer as it was received without any
disassemble and re-assemble overhead.  The reason why threads are used
is to make the locking mechanism efficient. This change boosted
the data rate to about 60\,MB/s per buffer machine.

\begin{figure*}[tb]
\centering
\includegraphics[width=135mm]{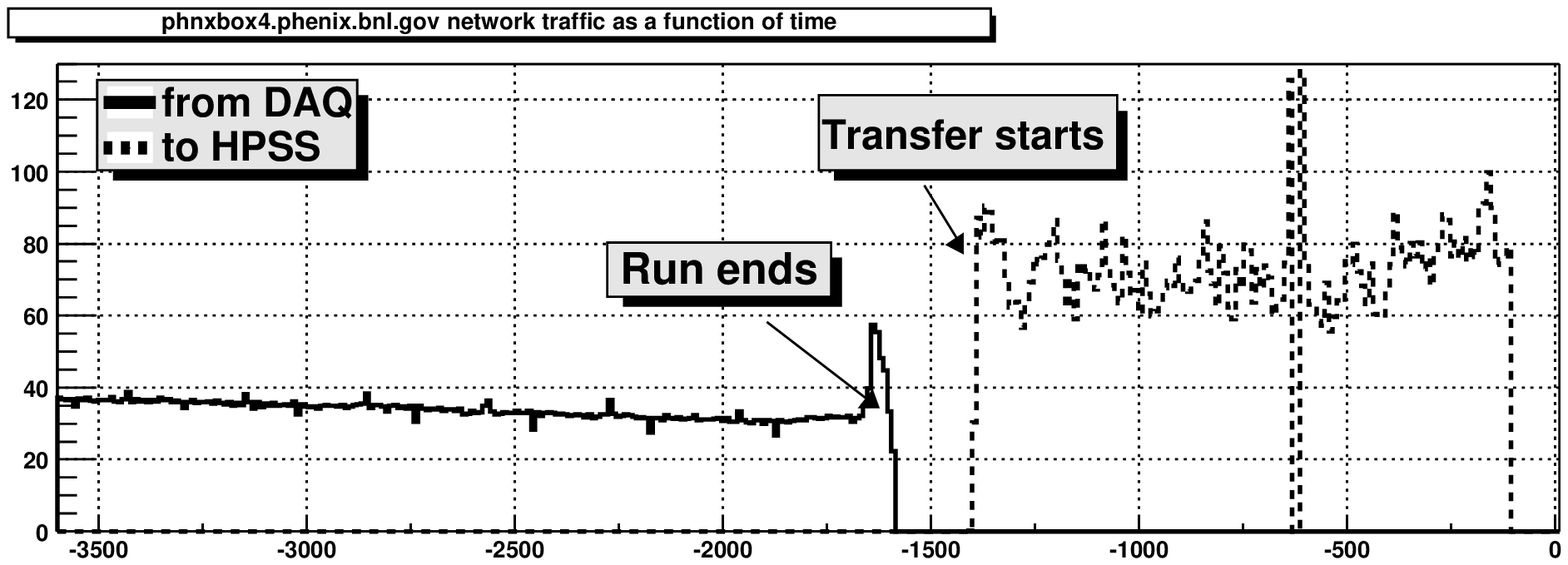}
\includegraphics[width=135mm]{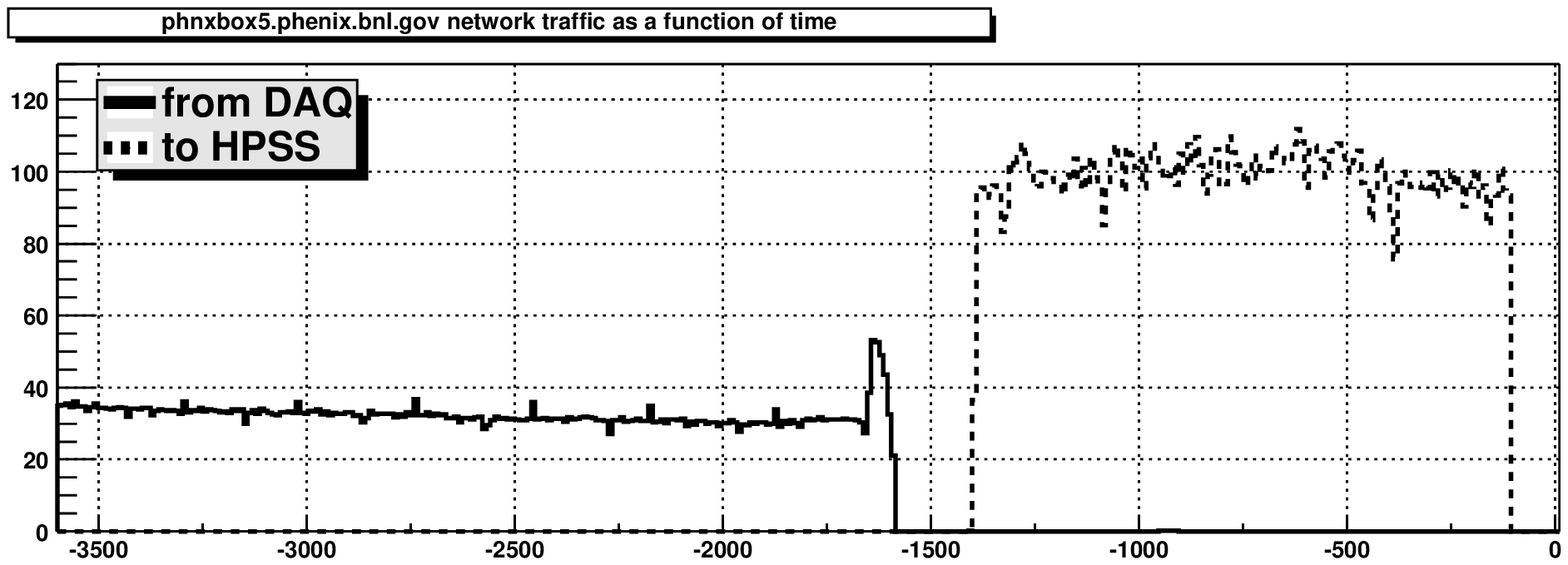}
\caption{\label{stripchart}
The data transfer rates measured for the two buffer machines. The
data represent 60 minutes of data. The snapshot shown here starts toward
the end of a RHIC fill. The solid line represents the data from the DAQ. 
After the beam is dumped, the transfer to HPSS starts (dashed curve).}
\end{figure*}

Fig.~\ref{stripchart} shows a strip chart of the data transfer rates
measured in the system. For the two buffer machines, we measure the
data throughput through the two network cards for both machines. The
data represent approximately 60 minutes of data. The snapshot shown
starts toward the end of a RHIC fill, and one can see the luminosity
decrease over the course of the first 30 minutes, giving modest data
rates of about 65\,MB/s with those beam conditions (solid line). When the
beam got dumped, our trigger detectors were sprayed with particles,
and the data rate surges to about 120\,MB/s for a short moment. We then
ended the run, and started the data transfer to HPSS (dashed line),
which runs at about 175MB/s for both machines combined.

\section{Online- and Near-line Monitoring} 

PHENIX has a successful tradition of using just one analysis
software package in online and offline. In this way, users have just
one learning curve, and there is only one set of libraries. In
offline and batch processing, the online capabilities are just not used.

We make use of the multi-threading capabilities of ROOT~\cite{ROOT} to
look at histograms and other information while data are being
processed by a background thread, which is required for online
monitoring.  This framework, called \emph{pMonitor}, is used for all
online monitoring tasks. It can read all supported types of data streams (file,
ET pools, and others), which is fully transparent to the user.  Other
packages (reconstruction, online calibrations, monitoring) build
on that basic framework.

For the data distribution among the various online monitoring
processes we use the ``ET'' system. The system consists of data pools,
into which processes can inject events, and other processes can
retrieve the events in various ways. The system supports a virtually
unlimited number of writers and readers. The ET system supports a
variety of access modes. Data can be tagged with additional
information (for example, the trigger type of that event) at the time
when they get inserted into the pool. Processes can then request only
events whose tags meet certain requirements and so request only events
they are truly interested in. Another access mode is ``shared
stream'', where a stream normally delivered to one consumer process is
split among two or more on a first-come, first-serve basis for load
balancing purposes. Normally, the same events get delivered to each
consumer.

Before the switch to the AML, a consumer process would request data
from the main ET pool where the data from the ATPs got merged, and
feed them into a secondary pool where all monitoring process read
from. With the AML, the ET system is no longer part of the logging
mechanism. However, the logger breaks the output stream up into files
of about 1.5\,GB, and at full speed, a new file is opened and the old
one closed approximately every 30 seconds. We now wait for a file to
be closed, and read a few hundred events from it and insert them into
the ET pool where the monitoring processes read from. In this way, we
did not need to change anything in the setup, and the only difference
is a time lag of about 30 seconds before the data reach the monitoring
processes.

A DAQ run in PHENIX is about one hour worth of data taking, which
yields a data volume of approximately 400GB at full speed, which gives
about 250 files of 1.5\,GB. Several quantities, such as calibration
constants, are valid for a whole run and usually need to be derived
from the data of the whole run. The time when the data are still in
the buffer disks is the last time when all run fragments are
conveniently together in the same place, and we use that opportunity to
analyze the data in various ways. 

One of the goals is to have all the required calibrations constants
for the event reconstruction ready by the time the data have to be
deleted from the buffer disks. This will reduce the load on the
offline computing farm and the tape robots because fewer data
retrieval operations will be needed, and allow a faster analysis of
the data. In addition, if the calibration constants can be derived
within hours, the Level-2 trigger algorithms can make use of them as
well.

Another goal is to extract events taken with certain triggers in
order to determine the trigger efficiencies. We write those events to
other files and analyze them in various ways.  For example, we fully
reconstruct events, determine the invariant mass of photon-like clusters
in the electromagnetic calorimeter, fit the peak corresponding to
$\pi^{\rm 0}$ mesons, and extract the mass from the fit parameters.
Obtaining the expected  $\pi^{\rm 0}$ mass is taken as a sign that the
current calibrations are roughly correct.

\section{Software Choices}

When making software choices, we had to take into account
the fact that the collaboration consists of more than 400 scientists
from many countries. Software in everyday use has to be freely
available to all collaborators, and must be free of export and other
restrictions. As much as possible, PHENIX uses Open-Source software or
software under the \emph{GNU Public License} (GPL), which makes
the software available to everyone. Also, open-source and GPL software
is virtually guaranteed to be available for the operating systems we
use, which is not always the case for proprietary software. 

There is only one proprietary software package in use that is needed
by most collaborators, the Objectivity Database system. A special
license arrangement makes the software available to all collaborators,
and the company has been forthcoming in resolving perceived licensing
conflicts. Still, the software is available only for a certain compiler
version, which imposes restrictions on operating system upgrades. We
will try to reduce the use of the software as much as possible in
favor of open-source database systems.

In the data acquisition system, two more proprietary packages are in
use. One is a CORBA implementation from Iona, \emph{ORBIX}, which is
used for all inter-process communications. The proprietary VxWorks
operating system is used in several front-end processors. Since the
software is not used outside the DAQ system, the availability is not
an issue, but there is concern about the market-driven
direction of the future development. 

\section{Data management}

With multiple-TB data sets, the management of the data becomes an
important aspect. Early on, we restricted the database to management
of meta-data, data about file locations, and properties, but not the raw
data itself. In routine running, we need to distribute files
automatically across file servers to balance the server load, and we
also need to avoid the replication of files and resolve ownership
issues (when can a file be deleted? When copied?). Also, there is the
aspect of replicating popular files and accessing the ``nearest
copy''.

We are using  a \emph{Data Carousel} interface to the
HPSS system that pools file requests and minimizes tape mounts. This gives
us a performance boost of about a factor of 7 over an unmanaged
first-come, first-serve model. On top of that we implemented a package
called \emph{ARGO}, which is a file catalog system that keeps track of
where files are located. We also added a name-based identification of
files which queries a database to locate a file. This system is called
\emph{FROG} (``File Replica On a Grid'') which has been integrated
into pmonitor. There are also APIs and Web interfaces to ARGO.

\section{Future Plans}

Several enhancements are planned for the Run 4, which will start in
fall of 2003.  We will replace the ATM switch in the Event builder
with a Gigabit switch, and increase the number and capacity of the
buffer machines.

Another planned enhancement has been made possible with the
AML. PHENIX has long had a raw data standard that includes a compressed
raw data format (which is \emph{not} a compressed data file, which
would need to get uncompressed before reading). A buffer that would
normally be written out to a file is instead compressed with the
gzip-compression algorithm. The resulting smaller sequence of data is
then put into another buffer whose header specifies that the payload
is a complete compressed buffer. This new, smaller buffer gets written
out instead. On readback, the I/O system recognizes the buffer header,
uncompresses the buffer into the original one, and passes this
uncompressed buffer on to the next software layer. In this way, the
compression is handled completely on a relatively low level of the I/O
system, and is transparent to the higher-level layers.

The compressed data format shrinks the size of the buffer to
approximately 45\% of the original buffer size. However, the
compression algorithm is very CPU time-consuming; on a fast CPU, one
second worth of data would need 3 seconds to get compressed. Before
the use of the AML, buffers to be written out were assembled in the
logger, and the CPU of the buffer machines would need to
perform that compression. This is much too slow and prevented the
use of the compressed data scheme in the data acquisition so far.

However, the AML now just receives and writes out ready-made buffers
from the ATPs, which could send already compressed buffers. In this
way, it is possible to distribute the load over many CPUs, and add
more processors as the need arises. For the next run we will try to
write compressed data routinely, which would essentially double the
disk capacity or the data rate, or a combination of both.

\section{Summary}

For the Run 3 of the RHIC accelerator, we have upgraded out data
acquisition system and the online computing infrastructure in several
ways.  We were able to boost the recorded data rate to about 120\,MB/s,
and perform online calibrations, event filtering, and online
reconstruction of selected events before the data leave the counting
house.

We have integrated a resource management system with our core analysis
packages, which will help to manage our resources better and increase
the throughput in the reconstruction processes.

For the next run, we will complete the transition to Gigabit networks
in all core DAQ components, and try to implement the routine
compressed data output.

\end{document}